\documentclass[preprint,12pt]{elsarticle}

\usepackage{amssymb}

\journal{Physics Letters B}

 \newcommand\beq{\begin{equation}}
 
 \newcommand\eeq{\end{equation}}
 \newcommand\beqn{\begin{eqnarray}}
 \newcommand\eeqn{\end{eqnarray}}
 \newcommand\GeV{{\rm GeV}}

\def\Re{\,{\rm Re}\,}
\def\Im{\,{\rm Im}\,}

\def\fm{\,\mbox{fm}}
\def\GeV{\,\mbox{GeV}}

\def\Pom{\mathbb{P}}
\def\Reg{\mathbb{R}}
\def\lsim{\mathrel{\rlap{\lower4pt\hbox{\hskip1pt$\sim$}}
    \raise1pt\hbox{$<$}}}
\def\gsim{\mathrel{\rlap{\lower4pt\hbox{\hskip1pt$\sim$}}
    \raise1pt\hbox{$>$}}}

\begin{document}

\begin{frontmatter}



\title{Probing the Pomeron spin structure with Coulomb-nuclear interference}


\author{B. Z. Kopeliovich$^1$}
\author{M. Krelina$^{2,3}$}
\author{I. K.~Potashnikova$^1$}

\address{\centerline{$^1$Departamento de F\'{\i}sica,
Universidad T\'ecnica Federico Santa Mar\'{\i}a }
{Avenida Espa\~na 1680, Valpara\'iso, Chile}}
\address{\centerline{$^2$FNSPE, Czech Technical University in Prague} 
{Brehova 7, 11519 Prague, Czech Republic}}
\address{\centerline{$^3$Physikalisches Institut, University of Heidelberg}
{Heidelberg 69120, Germany}}

\begin{abstract}

Polarized $pp$ elastic scattering at small angles in the Coulomb-nuclear interference (CNI) region offers a unique opportunity to study the spin structure of the Pomeron. 
Electromagnetic effects in elastic amplitude can be equivalently treated either as Coulomb corrections to the hadronic amplitude (Coulomb phase), or as absorption corrections to the Coulomb scattering amplitude. We perform the first calculation of the Coulomb phase for the spin-flip amplitude and found it significantly exceeding the widely used non-flip Coulomb phase.
The alternative description in terms of absorption corrections, though equivalent, turned out to be a more adequate approach for the Coulomb corrected spin-flip amplitude.  
Inspired by the recent high statistics measurements of single-spin asymmetry with the HJET polarimeter  at the BNL, we also performed a Regge analysis of data, aiming at disentangling the Pomeron contribution. However, in spite of an exceptional accuracy of the data, they do not allow to single out the Pomeron term, which strongly correlates with the major sub-leading Reggeons. A stable solution can be accessed only by making additional ad hoc assumptions, e.g. assuming  the Pomeron to be a simple Regge pole, or fixing some unknown parameters. Otherwise, in addition to the STAR data at $\sqrt{s}=200\GeV$ new measurements, say at $100\GeV$ or 
$500\GeV$, could become decisive.
\end{abstract}



\end{frontmatter}


\section{Introduction}

The Pomeron has been introduced in the Regge theory as a rightmost singularity in the complex angular momentum plane, having vacuum quantum numbers and dominating elastic scattering amplitude at high energies.
Originally, having no dynamical input, for the sake of simplicity, it has been assumed to be a Regge pole with the intercept $\alpha_\Pom(0)=1$, however later, the observed rise of the total cross sections with energy led to a higher value of the intercept $\alpha_{\Pom}(0)>1$ \cite{dubovikov}. Besides, the absorptive corrections, generating Regge cuts, make the structure of the singularity more complicated.

With the advent of QCD, it was realized that the Pomeron corresponds to gluonic exchanges in the $t$-channel, what naturally explains why the cross section is nearly constant, or slowly rising with energy, and why the elastic amplitude is predominantly imaginary. The spin structure of the Pomeron exchange amplitude is related to the helicity conserving quark-gluon vertex, this is why it has been widely believed that the Pomeron has no spin-flip component.

Experimental measurement of the hadronic spin-flip amplitude is a challenge.
Indeed, the single-spin asymmetry is proportional to $\sin(\Delta \phi)$,
where $\Delta \phi$ is the relative phase between spin-flip and non-flip amplitudes. If the Pomeron were a Regge pole, this phase shift would be exact zero.
Otherwise, it is expected to be small, suppressing spin effects in elastic $pp$ scattering.

A unique opportunity to get a sizable single-spin asymmetry $A_N$
is to arrange Coulomb-nuclear interference (CNI) between nearly real Coulomb and almost imaginary Pomeron elastic amplitudes.
In this case, the relative phase is optimal for single spin asymmetry.
Even if the Pomeron is spin-less, the Coulomb amplitude does have a known spin-flip
part, due to the existence of the anomalous magnetic moment of the proton, generating a considerable spin-flip amplitude. This was first proposed in \cite{kl}, and a peculiar  $t$-dependence of the single-spin asymmetry $A_N(t)$ was found (see also \cite{buttimore}) with a maximum of about $4.5\%$ at $t=t_{max}$ with
\beq
t_{max}=-\sqrt{3}\, \frac{8\pi\alpha_{em}}{\sigma_{tot}^{pp}}
\approx -0.0025\GeV^2,
\label{CNI}
\eeq
where $t$ and $s$ are the 4-momentum transfer squared and c.m. energy squared, respectively. 

If, however, the Pomeron also has a spin-flip component, the curve $A_N(t)$,
shifts up or down, keeping approximately the same shape, depending
on the phase and magnitude of the hadronic spin-flip. This was proposed in \cite{kz-spin} as a way to measure the Pomeron spin-flip.

The  present analysis of data on $A_N(t)$ covers a wide energy range, where dominance of the Pomeron term is not guaranteed. This is why we have to rely on Regge phenomenology including sub-leading Regge terms. The exceptionally high accuracy of the fix-target data give a chance to determine the spin-flip part of the  Pomeron amplitude.

\section{Spin structure of hadronic elastic amplitudes}

The elastic $pp$ amplitude is fully described by five independent helicity amplitudes $\phi_i(s,t)$ ($i=1,...5$) defined in \cite{Bilenky:1964pm, buttimore}.
The total and elastic cross sections and single-spin asymmetry $A_N(t)$ are expressed via these amplitudes,
\beqn
\sigma^{pp}_{tot} &=& 4\pi\, \textrm{Im} (\phi_1 +  \phi_3 )|_{t=0} \equiv
8\pi\, \textrm{Im}\phi_+(t=0),
\nonumber\\
  \frac{d \sigma^{pp}_{el}}{dt} &=& 2\pi \left\{ |\phi_1|^2 + |\phi_2|^2 + |\phi_3|^2 + |\phi_4|^2 + 4|\phi_5|^2 \right\},
\nonumber \\
  A_N \frac{d \sigma^{pp}_{el}}{dt} &=& -4\pi \,\textrm{Im} \left\{ (\phi_1 + \phi_2 + \phi_3 - \phi_4)\phi_5^* \right\}.
  \label{eq:pp:ANxsec}
\eeqn

The spin amplitudes $\phi_i$ contain the hadronic and electromagnetic parts, as well as their interferences.
In what follows we study all of them.

To simplify notations we neglect small amplitudes $\phi_{2,4}$ and relate the amplitudes $\phi_i$ to the spin-flip and non-flip elastic amplitudes,
\beqn
 \phi_+(q_T)  &=& f_{\mathbb{P}}^{nf}(q_T) + f_{\mathbb{R}}^{nf}(q_T);
    \nonumber
    \\
\phi_5(q_T) &=& f_{\mathbb{P}}^{sf}(q_T) + f_{\mathbb{R}}^{sf}(q_T) .
\label{phi-f}
\eeqn
Here we replaced the 4-momentum transfer squared by its transverse component squared, $t\equiv -q^2
\approx -q_T^2$. The longitudinal momentum transfer in elastic scattering is vanishingly small at high energies.

Relying on Regge phenomenology we single out two terms in the amplitudes (\ref{phi-f}). The first one, dominating at high energies, is usually called Pomeron.
Although the related singularity in the complex angular momentum plane is expected to have a complicated structure \cite{bfkl,dglap,book}, within a restricted energy range it 
can be treated as an effective Regge pole with an intercept above one. The intercept and the amplitude phase might be different for the non-flip and spin-flip components (e.g. see \cite{k-povh}), in contrast to a real Regge pole.

The second term represents the common contribution of the major sub-leading Reggeons ($f,\ \omega,\ \rho,\ a_2$) having highest intercepts $\alpha_{\mathbb{R}}(0)\approx0.5$, which we fix at this value. 
We have no tools to disentangle different kinds of Reggeons, because include in the analysis  only elastic $pp$ data. As long as we are hunting for the spin-flip part of the Pomeron amplitude $f_{\mathbb{P}}^{sf}$, which is presumably small, the Reggeons might be important even at high energies, because have large spin-flip component, especially iso-vector Reggeons $\rho$ and $a_2$.

\subsection{Non-flip amplitudes}

We parametrize the $s$ and small-$q_T$ dependences of the Pomeron and Reggeon amplitudes as,
\beqn
f_{\mathbb{P}}^{nf}(q_T) 
    &=&  h_{\mathbb{P}}^{nf}\,
 e^{-{1\over2}B_{\mathbb{P}} q_T^2};
    \left( \frac{s}{s_0} \right)^{\alpha_{\mathbb{P}}^{nf}(0)  -1} ;
\label{amplitudesP-nf}\\
f_{\mathbb{R}}^{nf}(q_T) 
    &=& h_{\mathbb{R}}^{nf}\, e^{-{1\over2}B_{\mathbb{R}}^{nf} q_T^2}\left( \frac{s}{s_0} \right)^{\alpha_{\mathbb{R}}^{nf}(0)  -1}.
\label{amplitudesR-nf}
\eeqn
We  fix the values of the intercepts, which are known \cite{pdg,landshoff} and are close to the values,
\begin{eqnarray}
\alpha_{\mathbb{P}}^{nf}(0) &=& 1.1;
\label{intercept-P}\\
\alpha_{\mathbb{R}}^{nf}(0) &=& 0.5.
\label{intercept-R}
\end{eqnarray}

The energy dependence of the $q_T$-slopes is assumed to be logarithmic, in accordance with the standard Regge-pole form,
\beqn
B_{\mathbb{P}}^{nf} &=& \left( B_{\mathbb{P}}^{0}\right)^{nf} + 2 \left(\alpha^\prime_{\mathbb{P}}\right)^{nf} \ln(s/s_0),
\label{slopesP-nf}\\
B_{\mathbb{R}}^{nf} &=& \left( B_{\mathbb{R}}^{0}\right)^{nf} + 2 \left(\alpha^\prime_{\mathbb{R}}\right)^{nf} \ln(s/s_0),
\label{slopesR-nf}
\eeqn
which is applicable even if the pole is effective, within a restricted energy interval.

The slope of meson Regge trajectories $\alpha^\prime_{\mathbb{R}}=0.9\GeV^{-2}$
is universal, because it is inversely proportional to the color-triplet string tension \cite{cnn}. The slope of the Pomeron trajectory is poorly known, since the glue-balls lying on this trajectory have not been well identified so far, and the trajectory $\alpha_{\mathbb{P}}(t)$ at negative $t$ is not linear \cite{peter}. Besides, proximity of the unitarity bound in high-energy partial elastic $pp$ amplitude, leads to a partial amplitude rising with energy only at large impact parameters. 
This experimental observation \cite{amaldi} once again emphasizes that the Pomeron is not a pole and the related parameters, $\alpha_{\mathbb{P}}(0)$, $\alpha^\prime_{\mathbb{P}}$, should be treated as effective values, which can be used only in a restricted energy range.
Therefore, the interaction radius and
the effective $\alpha^\prime_{\mathbb{P}}$ significantly increase in comparison with the bare Pomeron parameters \cite{k3p}, so it should be adjusted to data, as well as the constant $B_{\mathbb{P,R}}^{0}$.

We performed a two-parameter fit to data \cite{pdg} for $pp$ elastic slope with 
parametrizations (\ref{slopesP-nf}),
$s_0=1\GeV^2$, and found,
\beqn
 \left( B_{\mathbb{P}}^{0}\right)^{nf} &=& 8.67\pm 0.34\,\GeV^{-2};
 \nonumber\\
  \left(\alpha^\prime_{\mathbb{P}}\right)^{nf}  &=& 0.27\pm 0.02\,\GeV^{-2}.
\label{B-fit}
\eeqn
Notice that these slope values do not affect much our analysis of data 
 at small $|t|<0.02\GeV^2$.

The other parameters in the non-flip amplitudes Eqs.~(\ref{amplitudesP-nf}), (\ref{amplitudesR-nf}) were also fitted to data on total and differential elastic $pp$ cross section, and the ratio of real-to-imaginary parts of the forward elastic amplitude \cite{pdg}. Nonetheless, the real part of the Pomeron non-flip amplitude, was fixed by the derivative  relation obtained within the eikonal Regge model in \cite{gribov-m}, or with the general dispersion approach in \cite{bronzan-ks}. In the approximation of small $\alpha^{nf}_\Pom(0)-1$ this relation reads
\beq
\frac{{\rm Re}\,h_{\Pom}^{nf}(0)}
{{\rm Im}\,h_{\Pom}^{nf}(0)} =
\frac{\pi}{2}\,
\frac{\partial \ln [{\rm Im}\,f_{\Pom}^{nf}(0)]}{\partial\ln s},
\label{bronzan}
\eeq
where $f_{\Pom}^{nf}$ is given by (\ref{amplitudesP-nf}).

Notice that the simplified model of an effective Pomeron pole Eq.~(\ref{amplitudesP-nf}), we rely upon for the non-flip amplitude, fails at much higher energies of the LHC,
where data show the cross section rising much faster, as was predicted in \cite{k3p}. However, in the restricted energy range below $\sqrt{s}\leq 200\GeV$, we are interested in, the model of an effective Pomeron pole  describes data well \cite{landshoff}.
  
The sub-leading Reggeons are known to be subject to exchange degeneracy based on duality of the $t$- and $s$-channel descriptions for the amplitude. As a result, among the leading should Reggeons with intercepts $\alpha(0)\approx 0.5$ the pairs of  $f-\omega$ and $\rho-a_2$,  cancel in the imaginary,  but add up in real parts of the $pp$ (also $K^+p$) elastic amplitude. In reality such a symmetry is  broken, and 
a part of the Reggeon contribution shows up as falling  total $pp$ cross section at medium-high energies. We combine here all Reggeons in an effective one with intercept fixed at $\alpha^{nf}_{\Reg}(0)=0.5$, but unknown magnitude. Moreover, the  residue factor of such an effective Regge pole does not have the usual phase dictated by the value of $\alpha^{nf}_{\Reg}(0)$, so we fit $\Im h_{\mathbb{R}}^{nf}$ and $\Re h_{\mathbb{R}}^{nf}$ separately.

The  fit with the effective Pomeron and Reggeon poles includes 3 parameters,
\beqn
\Im h_{\mathbb{P}}^{nf}(0) &=& 1.89 \pm 0.002\GeV^{-2};
\nonumber\\
\Im h_{\mathbb{R}}^{nf}(0) &=& 10.25 \pm 0.053\GeV^{-2};
\nonumber\\
\Re h_{\mathbb{R}}^{nf}(0) &=& -11.69 \pm 0.417\GeV^{-2},
\label{45}
\eeqn
while  $\Re h_{\mathbb{P}}^{nf}(0)$ is determined by the relation (\ref{bronzan}).

These results for the non-flip amplitudes will be used in the further fit to data on single-spin asymmetry.

\subsection{Spin-flip amplitudes}

The spin-flip amplitudes are parametrized in analogy to Eqs.~(\ref{amplitudesP-nf}), (\ref{amplitudesR-nf})

\beqn
f_{\mathbb{P}}^{sf}(q_T) 
   &=& \frac{q_T}{m_N}\, h_{\mathbb{P}}^{sf}\, e^{-{1\over2}B_{\mathbb{P}}^{sf} q_T^2} \left( \frac{s}{s_0} \right)^{\alpha_{\mathbb{P}}^{sf}(0)  -1},
 \label{amplitudesP-sf}\\
f_{\mathbb{R}}^{sf}(q_T) 
   &=& \frac{q_T}{m_N}\, h_{\mathbb{R}}^{sf} e^{-{1\over2}B_{\mathbb{R}}^{sf} q_T^2} \left( \frac{s}{s_0} \right)^{\alpha_{\mathbb{R}}^{sf}(0)  -1} ,
   \label{amplitudesR-sf}
\eeqn
where
\beqn
B_{\mathbb{P}}^{sf} &=& \left( B_{\mathbb{P}}^{0}\right)^{sf} + 2 \left(\alpha^\prime_{\mathbb{P}}\right)^{sf} \ln(s/s_0);
\label{slopesP-sf}\\
B_{\mathbb{R}}^{sf} &=& \left( B_{\mathbb{R}}^{0}\right)^{sf} + 2 \left(\alpha^\prime_{\mathbb{R}}\right)^{sf} \ln(s/s_0).
\label{slopesR-sf}
\eeqn

None of the ingredients in Eqs.~(\ref{amplitudesP-sf})-(\ref{slopesR-sf}) are known, in particular, the Pomeron spin-flip amplitude, which is the main goal of the present study.

If the Reggeons were true Regge poles, the intercepts $\left(\alpha_{\mathbb{R}}(0)\right)^{sf}$ and slopes $\left(\alpha^\prime_{\mathbb{R}}\right)^{sf}$ should be the same for the spin-flip and non-flip amplitudes. In fact, that is quite a good approximation. The $\rho$-Reggeon trajectory has been well measured at positive and negative $t$, and is perfectly linear with the universal Regge slope. This shows smallness of corrections from  
the Regge-cuts, like $\rho$-${\mathbb{P}}$, which has nearly the same intercept $\alpha_{\rho{\mathbb{P}}}(0)=\alpha_{\rho}(0)+\alpha_{\mathbb{P}}(0)-1$. 
However the Regge slope is almost twice as small as for the $\rho$-pole.
Smallness of the Regge cut corrections is confirmed by calculation in the eikonal model.
Therefore, we fix $\left(\alpha^\prime_{\mathbb{R}}\right)^{sf}=0.9\GeV^{-2}$.

The first terms in Eqs.~(\ref{slopesP-sf}),(\ref{slopesR-sf}) are unknown, but should not be very different from the non-flip values, because are also controlled by the proton size. Moreover, the further analysis shows that the results are nearly independent of the $\left( B_{\mathbb{P,R}}^{0}\right)^{\!sf}$ values, because of smallness of $t$ in the analyzed data. The fit results hardly vary even if these spin-flip slopes are reduced down to zero. So a good approximation is to fix $\left( B_{\mathbb{P,R}}^{0}\right)^{\!sf}= \left( B_{\mathbb{P,R}}^{0}\right)^{\!nf}$.

\section{Coulomb amplitudes}

Small-angle single spin asymmetry $A_N$ is mainly due to interference of nearly real Coulomb spin-flip and almost imaginary non-flip hadronic amplitudes. Such a large phase difference is optimal according to  Eq.~(\ref{eq:pp:ANxsec}) to maximize $A_N$. Multiple electromagnetic interactions affect the phases of both Coulomb and hadronic amplitudes

While the magnitude of the hadronic spin-flip amplitude is still questionable \cite{buttimore}, the spin structure of the electromagnetic amplitude of $pp$ elastic scattering is well known. The Coulomb spin amplitudes (C) in impact parameter space have the eikonal form \cite{Kopeliovich:2000ez}, related to
the momentum representation by Fourier transformation,
\beqn
  \phi_+^{em}(q_T) &=&  \frac{i}{2\pi}\int d^2b\,e^{i\vec q_T\cdot\vec b}\,
\left(1 - e^{i\chi_C^{nf}(b)}\right),
  \label{55a}\\
  \phi_5^{em}(q_T) &=&\frac{1}{2\pi}\int d^2b\,e^{i\vec q_T\cdot\vec b}\,
\,\chi_C^{sf}(b)\,
 e^{i\chi_C^{nf}(b)},
   \label{55b}
\eeqn
with the non-flip and spin-flip eikonal phases,
 \beqn
\chi_C^{nf}(b) &=& - \frac{\alpha_{em}}{2\pi}\int d^2q_T\,
\frac{F^2_{1}(q_T^2)}{q_T^2+\lambda^2}\,
e^{-i\vec q_T\cdot\vec b};
\label{60a}\\
\chi_C^{sf}(b) &=&
-\frac{ \alpha_{em}\kappa_p}{4\pi m_p}\int d^2q_T\,
\frac{F_{1}(q_T^2)F_{2}(q_T^2)}{q_T^2+\lambda^2}\,
\frac{(\vec q_T\cdot\vec b)}{b}\,
e^{-i\vec q_T\cdot\vec b},
\label{60b}
\eeqn
respectively.
Here $\kappa_p=\mu_p-1=1.793$ is the anomalous magnetic moment of the proton.
$F_1(q_T^2)$ and $F_2(q_T^2)$ are the Dirac and Pauli electromagnetic formfactors, respectively.
They are related to the electric and magnetic fomfactors $(1+\gamma)F_1=G_E+\gamma G_M$; $(1+\gamma)\kappa\,F_2=G_M-G_E$, where $\gamma=q_T^2/4m_p^2$. At small $q_T^2\ll 4m_p^2$, we are interested in, we rely on the approximation $F_2\approx F_1$.

At small $q_T$ the formfactor
can be approximated by the Gaussian form,
\beq
    F_{1}(q_T) = e^{-\frac{1}{6}\langle r_{em}^2 \rangle_p\, q_T^2},
    \label{eq:pp:GE-1}
\eeq
where $\langle r_{em}^2 \rangle_p$ is the proton mean charge radius squared.
We fix it at the value $\sqrt{\langle r_{em}^2 \rangle_p}=0.875\fm$  \cite{pdg}.

Notice that in the parametrization proposed in \cite{buttimore}, and used in all following data analyses, the slopes of elastic $pp$ scattering
and of the electromagnetic formfactor were taken equal, which is an oversimplification. One of them, the hadronic slope $B_{pp}(s)$, rises with energy, while another one, in Eq.~(\ref{eq:pp:GE-1}), is energy independent. We rely on the more realistic parametrization, explained above.

In order to keep the integrals in Eqs.~(\ref{60a}) and (\ref{60b}) finite we supply the photon with a small mass
$\lambda$ which disappears from the final expressions.
Notice that the pure Coulomb amplitude  has a nonzero phase coming from the higher order
terms in (\ref{55a})-(\ref{55b}), e.g. two photon exchange amplitude is imaginary.

\section{Coulomb-nuclear interference}

\begin{figure}[!htb]
  \includegraphics[scale=0.3]{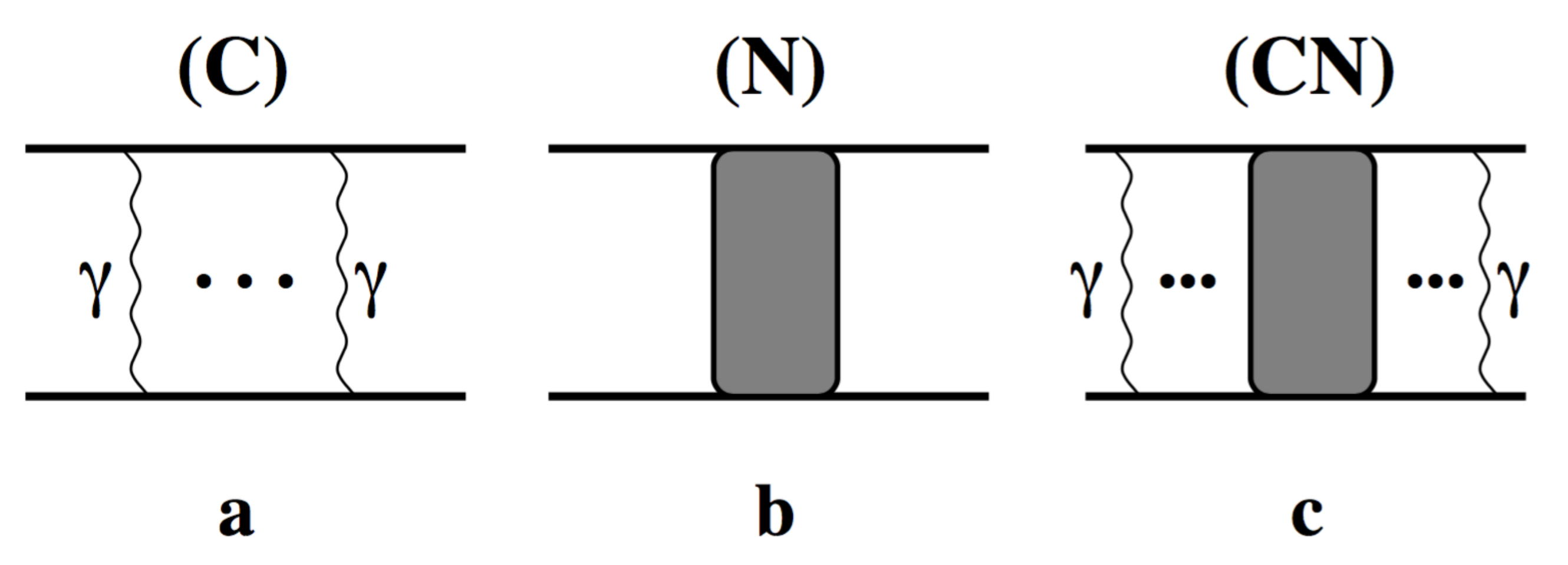}\centering
  \caption{Three types of interaction: pure electromagnetic ({\bf a}),
pure strong interaction ({\bf b}), and combined strong and electromagnetic interactions ({\bf c}). }
  \label{fig:graphs}
\end{figure}

The long-range Coulomb forces also affect the strong-interaction amplitude. This is illustrated in Fig.~\ref{fig:graphs}, following the consideration of this problem in \cite{Kopeliovich:2000ez}.
These graphs can be grouped and interpreted differently. One way, employed in \cite{Kopeliovich:2000ez}, is to combine the last two graphs, (N) and (CN), and treat it as a Coulomb modified strong-interaction amplitude. The modification is approximated by giving an extra phase factor to the hadronic amplitude. This factor is called in the literature Coulomb phase \cite{bethe,cahn,Kopeliovich:2000ez}.
 
\subsection{Coulomb phase shift of hadronic amplitudes}

This effect has been calculated so far \cite{bethe,cahn,Kopeliovich:2000ez} only for non-flip amplitudes, but applied incorrectly to the spin effects. Here we derive the Coulomb-modified phases for all spin amplitudes.

Using (\ref{60a}) and (\ref{60b}) 
we can calculate the phases of non-flip and spin-flip electromagnetic amplitudes,
Eq.~(\ref{55a}) and (\ref{55b}) respectively, as,
\beqn
\delta_C^{nf}(q_T) &=&\frac{2\pi\,\phi_+^{em}(q_T)}{\int d^2b\,
e^{i\vec q_T\cdot\vec b}\,
\chi_C^{nf}(b)}-1;
\label{eq:delta-nf}
\\
\delta_C^{sf}(q_T) &=&\frac{2\pi\,\phi_5^{em}(q_T)}{\int d^2b\,
e^{i\vec q_T\cdot\vec b}\,
\chi_C^{sf}(b)}-1.
\label{eq:delta-sf}
\eeqn

If the multiple Coulomb interactions generated the same phase shift for the electromagnetic  (C) and hadronic (N+CN) amplitudes, there would be no effect on the spin-asymmetry $A_N$ at all.
However, the Coulomb induced phase shifts of the two term (C) and (N+CN) depicted in Fig.~\ref{fig:graphs} are different and the difference is usually called Coulomb phase.

The non-flip phase Eq.~(\ref{eq:delta-nf}) was calculated in \cite{Kopeliovich:2000ez} analytically and our numerical calculations confirm that result. The spin-flip phase
Eq.~(\ref{eq:delta-sf}) is calculated here for the first time.

The hadronic part of the amplitudes includes the
two other terms in Fig.~\ref{fig:graphs} combined together, (N)+(NC), which correspond to the the contribution of strong interactions Eq.~(\ref{phi-f}), modified by Coulomb corrections. The non-flip amplitude reads \cite{Kopeliovich:2000ez},
\beq
\left.\phi_+(s,q_T)\right|_{(N)+(NC)}= \frac{i}{2\pi}\int d^2b\,
e^{i\vec q_T\cdot\vec b}\,
e^{i\chi_C^{nf}(b)}\, \gamma_N^{nf}(b),
\label{30}
\eeq
where
\beq
\gamma_N^{nf}(b)=\frac{i}{2\pi}\int d^2q_T\,e^{-i\vec q_T\cdot\vec b}\,
\phi_+(q_T),
\label{40}   
\eeq
and $\phi_+(q_T)$ is given by Eq.~(\ref{phi-f}). The phase of this amplitude is given by,
\beq
    \left.\delta^{nf}(q_T)\right|_{(N)+(NC)}= \frac{\left.\phi_+^h(q_T)\right|_{(N)+(NC)}}
{\left.\phi_+^h(q_T)\right|_{(N)}}
 -1.
\label{44}
\eeq
We assume here that the phase is small, $\delta\ll1$, which is justified by the higher order ($\alpha_{em}^2$) corrections, related to the second and higher terms in the expansion of the exponential $\exp(i\chi_C^{nf})$ in Eq.~(\ref{55a}). The smallness of the Coulomb correction allows to represent it as a small shift of the phase.

Notice that both terms, $(N)$ and $(NC)$, contributing to (\ref{30}), are controlled by short-range strong interactions and have a sizeable magnitude only at small impact parameters, $b^2\lesssim 2B_{pp}$. The Coulomb forces nevertheless, considerably affect the phase of the combined amplitude. The Coulomb phase for the non-flip hadronic amplitude is given by the difference between (\ref{44}) and (\ref{eq:delta-nf}).

The spin-flip amplitude has a structure, analogous to Eq.~(\ref{30}), but the hadronic non-flip factor $\gamma_N^{nf}(b)$ should be replaced by a spin-flip amplitude, either hadronic, or Coulomb. So we get,
\beq
\left.\phi_5(s,q_T)\right|_{(N)+(NC)}= \frac{i}{2\pi}\int d^2b\,
e^{i\vec q_T\cdot\vec b}\, e^{i\chi_C^{nf}(b)}
\left[\chi_C^{sf}(b)\, \gamma_N^{nf}(b)+
\gamma_N^{sf}(b)\right].
\label{50}
\eeq
The first term here is given by Eqs.~(\ref{60b}) and (\ref{40}).
The second term is given by the Fourier transformed hadronic spin-flip amplitude Eq.~(\ref{phi-f}).

So far, the non-flip Coulomb phase shift Eq.~(\ref{44}) has been used for the hadronic spin-flip
amplitude \cite{buttimore,trueman,star,HJET-PRL}. The Coulomb corrected hadronic spin-flip amplitude is given by Eq.~(\ref{50}). However converting  it to a Coulomb phase shift might be problematic.
As was mentioned above, the relative value of the Coulomb corrections in the non-flip amplitude, Eq.~(\ref{45}) is suppressed by $\alpha_{em}$. However, the relative
magnitude of the Coulomb correction in the spin-flip amplitude, given by the first term in Eq.~(\ref{50}), is much larger, of the order of $\alpha_{em}h_{\mathbb{P}}^{nf}/h_{\mathbb{P}}^{sf}$. 

Another source of enhancement of the Coulomb correction is the less singular behavior of the spin-flip amplitude, $1/q_T$, compared with the quadratic singularity, $1/q_T^2$, in the non-flip amplitude. So the spin-flip  Coulomb interaction is less peripheral, and is more  affected by the interference with short-range strong interactions.

Such a large Coulomb correction cannot be represented as a phase shift, because it also affects the absolute value of the amplitude. Therefore, in the next section, we re-group the graphs in Fig.~\ref{fig:graphs} in a way that the modification acquires a meaning of hadronic corrections to the Coulomb spin-flip amplitude. Such a more accurate calculation of Coulomb-nuclear interference is used for further analysis of data.

\subsection{Absorptive corrections}

One can group the graphs in Fig.~\ref{fig:graphs}
differently, so that the result can be interpreted as absorption corrections to the Coulomb amplitude. Of course, the final results must remain unchanged, either for the spin-flip, or non-flip amplitudes, and  numerical comparison confirms that.

In all calculations of the CNI contribution to single-spin asymmetry, performed so far \cite{buttimore,trueman,star,HJET-PRL}, the Coulomb phase applied to the hadronic spin-flip amplitude, has been taken from spin non-flip calculations \cite{bethe,cahn,Kopeliovich:2000ez}. Such a procedure is unjustified, moreover, is quite incorrect, as was demonstrated in the previous section.

If one  combines the graphs (C) and (CN) depicted in Figs.~\ref{fig:graphs}a and \ref{fig:graphs}c respectively, one gets Coulomb amplitude with absorption corrections related to  possibility of strong inelastic interactions, destroying the rapidity gap. This is why it is also called the amplitude of survival probability of a large rapidity gap, associated with elastic Coulomb interaction of hadrons.

Absorption corrections are most effectively calculated in impact parameter representation. One should Fourier transform the $q_T$-dependent electromagnetic amplitudes to $b$-space, like was done in Eqs.~(\ref{55a}), (\ref{55b}).
Then introduce the absorptive factor,
\beq
\phi^{em}(b)\Rightarrow \phi^{em}(b)\times S(b),
\label{factor-S}
\eeq
where
\beq
S(b)=1+2i\gamma^{nf}_N(b),
\label{S}
\eeq
and $\gamma^{nf}_N(b)$ is defined in (\ref{40}). To avoid a terminological confusion, notice that the correction, corresponding to the graph  in Fig.~\ref{fig:graphs}c, is not pure absorptive, i.e. imaginary, but $\gamma^{nf}_N(b)$ contains a small real part.

Now we are in a position to calculate the absorption corrected $q_T$-dependent electromagnetic helicity amplitudes by making inverse Fourier transformation to momentum representation,
\beqn
  \widetilde\phi_+^{em}(q_T) &=&  \frac{i}{2\pi}\int d^2b\,e^{i\vec q_T\cdot\vec b}\,
\left(1 - e^{i\chi_C^{nf}(b)}\right)\,S(b)
\nonumber\\ &=&
\phi_+^{em}(q_T) -  \frac{i}{2\pi}\int d^2b\,e^{i\vec q_T\cdot\vec b}\,
\left(1 - e^{i\chi_C^{nf}(b)}\right)\,[1-S(b)],
\label{70a}
\eeqn
where $\chi_C^{nf}(b)$ is given by (\ref{60a}).
The absorptive correction here is given by the second term here, which does not contain a long-range divergency, because the factor $1-S(b)$ vanishes at $b^2\gg B_{pp}$.


The absorption corrected spin-flip electromagnetic amplitude Eq.~(\ref{55b}) has analogous form,
\beqn
  \widetilde\phi_5^{em}(q_T) &=&  \frac{i}{2\pi}\int d^2b\,e^{i\vec q_T\cdot\vec b}\,
\,\chi_C^{sf}(b)\, e^{i\chi_C^{nf}(b)}
\,S(b) \nonumber\\
&=&  \phi_5^{em}(q_T)-\frac{i}{2\pi}\int d^2b\,e^{i\vec q_T\cdot\vec b}\,
\,\chi_C^{sf}(b)\,
 e^{i\chi_C^{nf}(b)}
\,[1-S(b)].
\label{70b}
\eeqn

Within the approach used in this Sect.~4.2 the amplitude 
$\tilde{\phi}^{em}_5$ (l.h.s. of (\ref{70b})) is given by the sum of Figs.~\ref{fig:graphs}a and \ref{fig:graphs}c terms. Thus it should contain also the Pomeron spin-flip contribution $\gamma^{sf}_N$ (similar to the second term in r.h.s. of (\ref{50})) modified by the Coulomb phase $\exp(i\chi^{nf}_C(b))$.
 However here we consider the small $q_T$ region where the asymmetry
caused by the interference of the Pomeron spin-flip amplitude Fig.~\ref{fig:graphs}b with the Coulomb non-flip amplitude $\phi^{em}_+$ is enhanced by the   
singular $1/t$ factor in Coulomb amplitude, while the $\gamma^{sf}_N$ contribution from  $\tilde{\phi}^{em}_5$ (\ref{70b}) is enhanced by the 
$\ln(1/t)$ only. Therefore, in comparison with a much larger,
$\phi^{em}_+\times  f^{sf}_P$ contribution, we neglect this term in (\ref{70b}) and in our further analysis.

\section{Data analysis: spin-flip amplitudes}

Now we can calculate the single-spin asymmetry $A_N(t)$, Eq.~(\ref{eq:pp:ANxsec}), since all the spin-flip amplitudes
(\ref{amplitudesP-sf}), (\ref{amplitudesR-sf}), (\ref{70b}), and non-flip amplitudes (\ref{amplitudesP-nf}), (\ref{amplitudesR-nf})(\ref{70a}),
amplitudes are either known, or parametrized. 

At small $t$  the dominant contribution to $A_N$ comes
from the interference of hadronic and electromagnetic amplitudes.
At  high energies the former is expected to be nearly imaginary, while the latter is almost real. Such a large phase shift allows to maximize the single-spin asymmetry $A_N$\cite{kl}. Of course, at medium-high energies, the sub-leading Reggeons  with intercepts $\alpha_{\mathbb{R}}(0)\approx 0.5$ can supply a considerable real part. Besides, the iso-vector Reggeons ($\rho, a_2$) are predominantly spin-flip, so contribute to the hadronic spin-flip amplitudes, which can interfere with electromagnetic non-flip component.

We performed a fit simultaneously to all available data for $A_N(q_T)$, with 5 unknown parameters, the real and imaginary parts of the spin-flip Pomeron and Reggeon amplitudes in Eq.~(\ref{amplitudesP-sf}), (\ref{amplitudesR-sf}). The fifth parameter is the Pomeron spin-flip intercept $\alpha_{\mathbb{P}}^{sf}(0)$, which might be different from the known non-flip value. If the Pomeron was a Regge pole, the spin-flip and non-flip intercepts would coincide. However, none of the contemporary dynamic models for the Pomeron support its Regge pole origin, some even predict a considerably  larger value of $\alpha_{\mathbb{P}}^{sf}(0)$ \cite{k-povh}.

\begin{figure}[!htb]
  \includegraphics[scale=0.5]{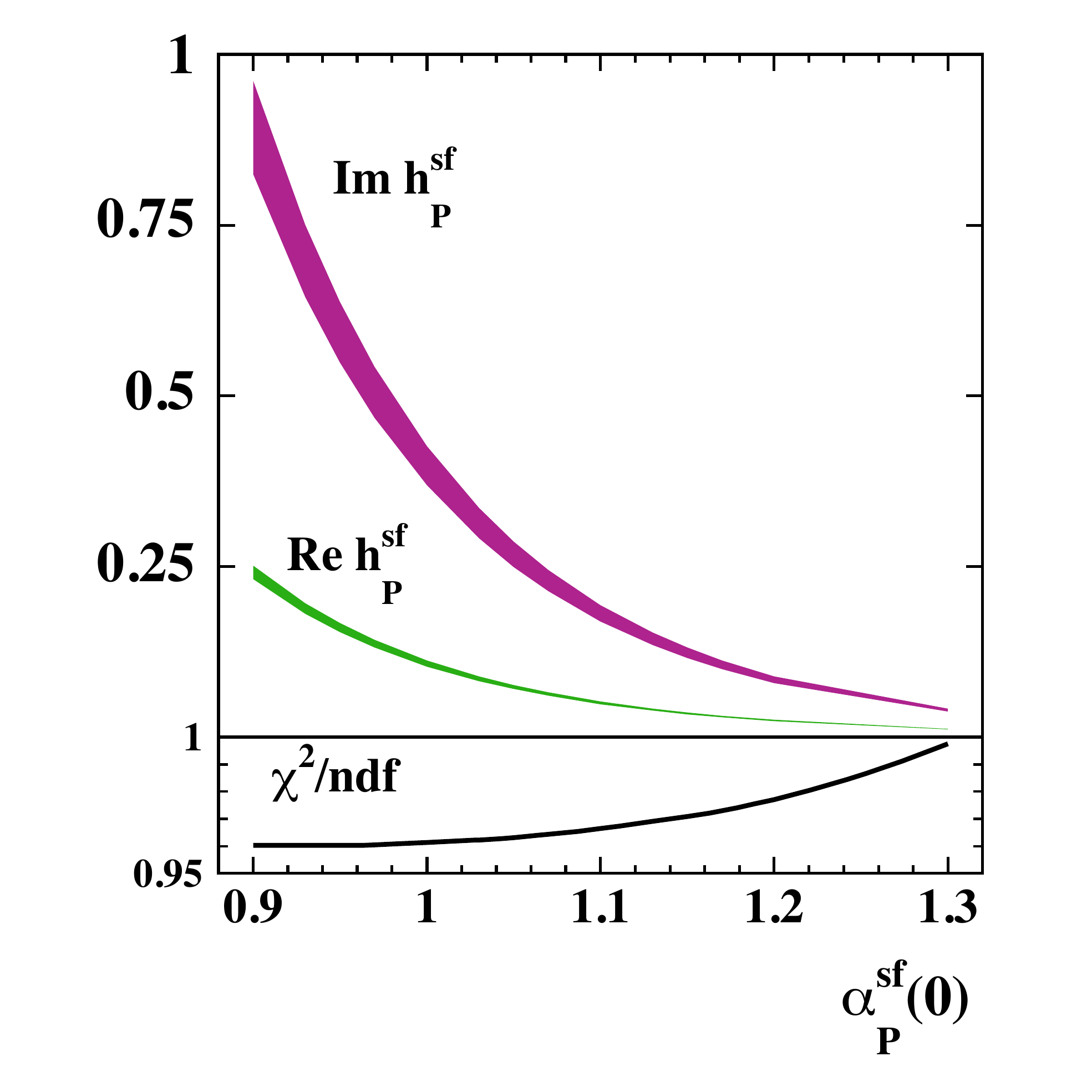}\centering
  \caption{(Color online) {\sl Upper panel:} real and imaginary parts of the factor $h_{\mathbb{P}}^{sf}$,
  which defines the magnitude of the Pomeron spin-flip component defined in (\ref{amplitudesP-sf}), (\ref{amplitudesR-sf}) vs the effective intercept $\alpha_{\mathbb{P}}^{sf}(0)$.
 {\sl Bottom panel:} $\chi^2/ndf$ vs $\alpha_{\mathbb{P}}^{sf}(0)$. }
  \label{fig:alpha}
\end{figure}

Our fit to data \cite{star,HJET-PRL,704} revealed a strong correlation between
$\alpha_{\mathbb{P}}^{sf}(0)$ and other parameters. The $\chi^2$ profile of this parameter is plotted in  Fig.~\ref{fig:alpha}. 
For each fixed value of $\alpha_{\mathbb{P}}^{sf}(0)$ other four parameters are fitted and their values strongly correlate with the chosen value of $\alpha_{\mathbb{P}}^{sf}(0)$. 
One can see that the value of $\chi^2$ is close to the number of degrees of freedom (ndf) within a wide range of $\alpha_{\mathbb{P}}^{sf}(0)=0.9-1.3$.
This shows that preferable value of $\alpha_{\mathbb{P}}^{sf}(0)$ cannot be reliably localized because all high statistics data are available only at medium-high energies, where Reggeon contribution is large and strongly correlates with the Pomeron. The data at $\sqrt{s}=200\GeV$ alone, of course cannot determine $\alpha_{\mathbb{P}}^{sf}(0)$, only together with lower energy HJET data \cite{HJET-PRL}, which suffer of strong correlations. Apparently new data at collider energies are required.
 
The real and imaginary parts of the Pomeron spin-flip are plotted with their error band in Fig.~\ref{fig:alpha} vs the fixed value of $\alpha_{\mathbb{P}}^{sf}(0)$. Interestingly, the spin-flip Pomeron amplitude turns out to be predominantly imaginary, like the non-flip one. However, the real-to-imaginary ratio is much different from what the differential relation (\ref{bronzan}) would give, if were naively applied to the spin-flip amplitude (where it has never been proven). In such a case it would be negative at $\alpha_{\mathbb{P}}^{sf}(0)<1$

We conclude that unfortunately  energy dependence of the spin-flip Pomeron amplitude cannot be determined from available data because of large correlations with other Reggeons at low energies.
 
To show an example of full set of other parameters we  choose the value of $\alpha_{\mathbb{P}}^{sf}(0)=1.1$, the same as for non-flip (\ref{amplitudesP-nf}).
\beqn
    \alpha_{\mathbb{P}}^{sf}(0) &=& 1.1 ({\rm fixed})
    \nonumber\\
    \Im h_{\mathbb{P}}^{sf} &=&   0.177   \pm   0.0122 \GeV^{-2}
    \nonumber\\
\Re h_{\mathbb{P}}^{sf} &=&   0.048   \pm   0.002 \GeV^{-2}
 \nonumber\\
   \Im h_{\mathbb{R}}^{sf} &=&  -4.352   \pm   0.370 \GeV^{-2}
 \nonumber\\
   \Re h_{\mathbb{R}}^{sf} &=&  -2.233   \pm   0.064 \GeV^{-2}
  \label{80c}\\
   \chi^2/ndf  &=&    315.9/329
 \nonumber
 \eeqn
The choice of  $\alpha_{\mathbb{P}}^{sf}(0)=\alpha_{\mathbb{P}}^{nf}(0)$ is made just for convenience, to make the fractional spin-flip of the Pomeron, $ r_5^{\mathbb{P}}$ \cite{buttimore} independent of energy,
\beq
r_5^{\mathbb{P}} =
\frac{m_N\, f_{\mathbb{P}}^{sf}(q_T)}
{q_T\Im f_{\mathbb{P}}^{sf}(q_T) }
\label{r5}
\eeq
 For the above sample of parameters Eq.~(\ref{80c}) $r_5$ is energy independent,
\beq
\Im r_5=0.094 \pm 0.006.
\label{r5-fit}
\eeq
This ratio also can be treated as the anomalous magnetic moment of the Pomeron 
$\mu_{\mathbb{P}} = 2 r_5^{\mathbb{P}}$,  introduced in \cite{kz-spin}.

\begin{figure}[!htb]
  \includegraphics[scale=0.8]{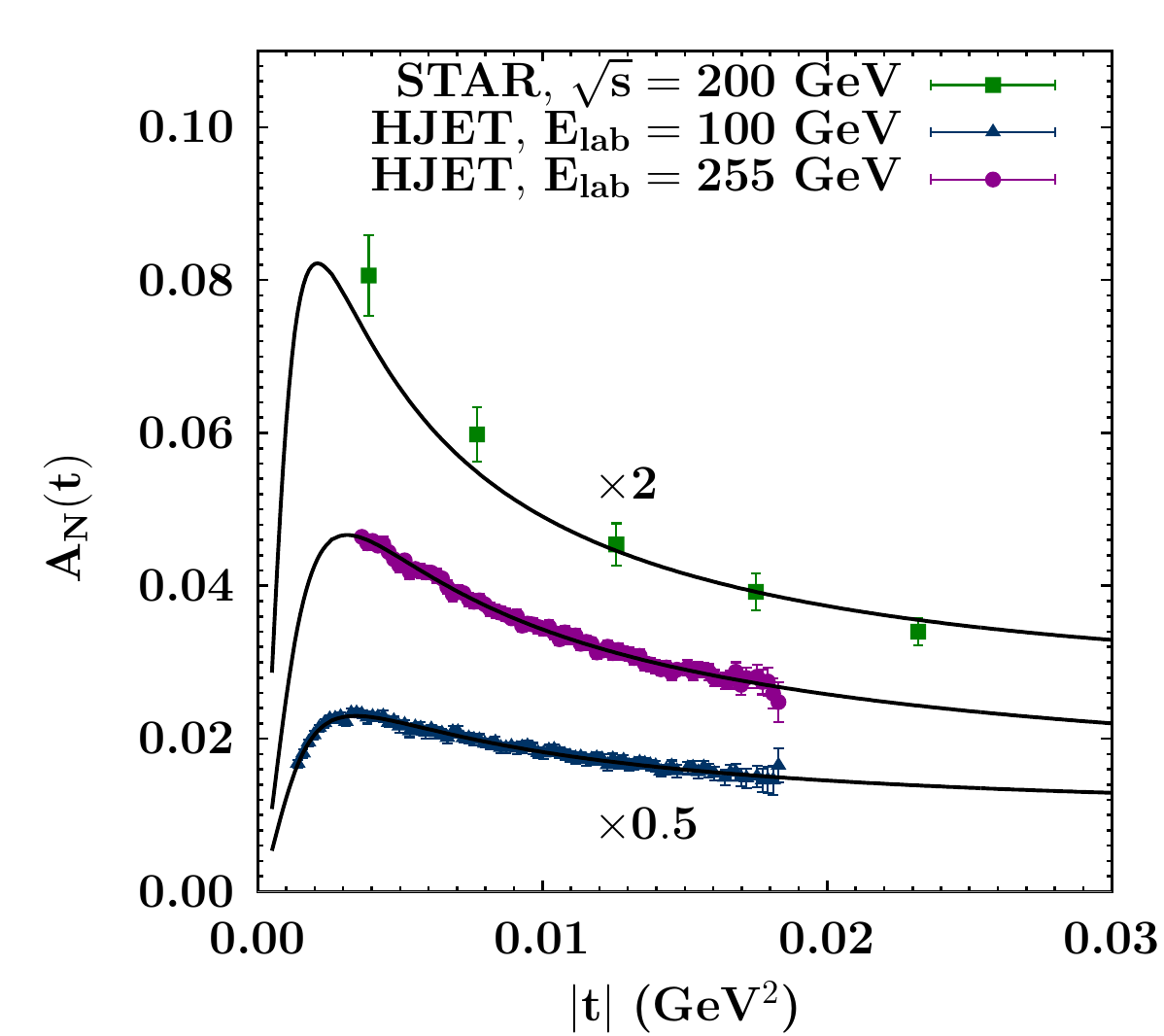}\centering
  \caption{(Color online) Collider RHIC  \cite{star}  and fixed-target \cite{HJET-PRL} data
  vs calculations with
  the parameters (\ref{80c}). }
  \label{fig:data}
\end{figure}

As was mentioned, Fig.~\ref{fig:alpha} also shows a good description of data with $\chi^2\approx ndf$ for a wide range of effective intercepts. To visualize the quality of description we plotted
$A_N(t)$ calculated with the parameters (\ref{80c}) in comparison with STAR \cite{star} and HJET data in Fig.~\ref{fig:data}.  

\section{Summary and discussion}

We analyzed data on single-spin asymmetry $A_N(t)$ in small-angle $pp$ elastic scattering, where it is 
presumably dominated by Coulomb-nuclear interference. The main objective of our analysis was the Pomeron spin structure and its energy dependence.

Electromagnetic corrections to the hadronic amplitude, widely known as Coulomb phase, have never been derived for the spin-flip component of the hadronic amplitude, but the non-flip Coulomb phase has been incorrectly applied to the spin-flip amplitude. We calculated the spin-flip Coulomb phase and found it significantly exceeding that for the non-flip amplitude. Moreover, typically the corrections are so large, that hardly can be treated as a phase shift.

Electromagnetic corrections can be equivalently interpreted either as Coulomb corrections to the hadronic amplitude, or as hadronic (absorption) corrections to the Coulomb amplitude. Although the latter interpretation is more traditional, the former is more adequate for the spin-flip amplitude, and offers easier evaluation. 

The wide energy range of currently available data allows to perform a Regge analysis of
the spin-flip hadronic amplitudes, aiming at disentangling the Pomeron and Reggeon terms.
However, in spite of high statistics of data from the fixed-target HJET measurements, these two contributions in the spin-flip amplitude cannot be reliably separated because of strong correlations.
Fig.~\ref{fig:alpha} demonstrates that the results strongly depend on the value of the unknown effective intercept $\alpha_{\mathbb{P}}^{sf}(0)$.
The Pomeron is not a Regge pole, so the effective intercept of its spin-flip and non-flip components might be quite different (e.g. see \cite{k-povh}). 

Unfortunately, data at sufficiently high energies,  to neglect the Reggeon contribution, are available only at one energy $\sqrt{s}=200\GeV$, what is insufficient for determination of $\alpha_{\mathbb{P}}^{sf}(0)$. One can fit data well with about the same quality, fixing $\alpha_{\mathbb{P}}^{sf}(0)$ at different values as is demonstrated in Fig.~\ref{fig:alpha}. At the same time the real and imaginary parts of the Pomeron spin-flip amplitude vary considerably.
Additional measurements at a different collider energy $\sqrt{s}$, e.g. $100\GeV$ or $500\GeV$ could solve the problem. 

The recently published alternative Regge analysis \cite{HJET-PRL} of the same data deserves commenting, since it arrived at quite different conclusions. The reason is the additional unjustified assumptions made in the analyses. The Pomeron was assumed to be a "simple" Regge pole, contradicting any  theoretical expectation (see e.g. \cite{bfkl,dglap,book}).  Fig.~\ref{fig:alpha}
demonstrates that just one ad hoc assumption about the value of $\alpha_{\mathbb{P}}^{sf}(0)$ immediately leads to certain fit results with small errors.

\section*{Acknowledgements}

We are thankful to Andrei Poblaguev and Wlodek Guryn for informative discussions.
The work of B.Z.K. and I.K.P. was supported in part by grants ANID - Chile FONDECYT 1170319, and by ANID PIA/APOYO AFB180002.\\
The work of M.K. was supported by the project Centre of Advanced Applied Sciences CZ.T02.1.01/0.0/0.0/16-019/0000778 and by International Mobility of Researchers - MSCA IF IV at CTU in Prague \\CZ.02.2.69/0.0/0.0/20\_079/0017983, Czech Republic.
It was also supported at the initial stage by the CONICYT Postdoctorado N.3180085 (Fondecyt Chile).

\end{document}